\documentclass{emulateapj}
\usepackage{graphicx}
\usepackage[space]{grffile}
\usepackage{latexsym}
\usepackage{textcomp}
\usepackage{longtable}
\usepackage{multirow,booktabs}
\usepackage{amsfonts,amsmath,amssymb}
\usepackage{natbib,verbatim}
\usepackage{url}
\usepackage{hyperref}
\hypersetup{colorlinks=true,linkcolor=red,urlcolor=purple,citecolor=blue,pdfborder={0 0 0}}
\newif\iflatexml\latexmlfalse
\usepackage[utf8]{inputenc}
\usepackage[english]{babel}
\usepackage{todonotes}
\usepackage{xspace}

\newcommand{\msun}{\,\mathrm{M}_\odot}

\newcommand{\myr}{\,\mathrm{Myr}} 
\newcommand{\pc}{\,\mathrm{pc}}
\newcommand{\au}{\,\mathrm{au}} 
\newcommand{\rsun}{\,\mathrm{R}_\odot}
\newcommand{\Bex}{{B$_\mathrm{ex}$}\xspace}
\newcommand{\Aex}{{A$_\mathrm{ex}$}\xspace}

\shorttitle{Insights on the origin of the S-stars in the Galactic center}
\shortauthors{Trani et al.}

\begin{document}


\title{The Keplerian three-body encounter I.\\ Insights on the origin of the S-stars and the G-objects in the Galactic center}


\author{Alessandro A. Trani\altaffilmark{1,$\star$,$\dagger$}}
\author{Michiko S. Fujii\altaffilmark{1}}
\author{Mario Spera\altaffilmark{2,3,4,5,6}}

\altaffiltext{1}{Department of Astronomy, Graduate School of Science, The University of Tokyo, 7-3-1 Hongo, Bunkyo-ku, Tokyo, 113-0033, Japan}
\altaffiltext{2}{INAF, Osservatorio Astronomico di Padova, Vicolo dell'Osservatorio 5, I--35122, Padova, Italy}
\altaffiltext{3}{INFN, Sezione di Padova, Via Marzolo 8, I–35131, Padova, Italy}
\altaffiltext{4}{Dipartimento di Fisica e Astronomia ‘G. Galilei’, University of Padova, Vicolo dell’Osservatorio 3, I–35122, Padova, Italy}
\altaffiltext{5}{Department of Physics and Astronomy, Northwestern University, Evanston, IL 60208, USA}
\altaffiltext{6}{Center for Interdisciplinary Exploration and Research in Astrophysics (CIERA), Evanston, IL 60208, USA}

\altaffiltext{$\star$}{Email: aatrani@gmail.com}
\altaffiltext{$\dagger$}{JSPS Fellow}




\begin{abstract}
Recent spectroscopic analysis has set an upper limit to the age of the S-stars, the ${\sim}30$ B-type stars in highly eccentric orbits around the supermassive black hole (SMBH) in the Galactic center. { The inferred age (${<}15\myr$) is in tension with the binary break-up scenario proposed to explain their origin. However, the new estimate is compatible with the age of the disk of O-type stars that lies at a farther distance from the SMBH.} 
Here we investigate a new formation scenario, assuming that both S-stars and the O-type stars were born in the same disk around SgrA*. 
We simulate encounters between binaries of the stellar disk and stellar black holes from a dark cusp around SgrA*.
{ We find that B-type binaries can be easily broken up by the encounters and their binary components are kicked into highly eccentric orbits} around the SMBH. In contrast, O-type binaries are less frequently disrupted and their members remain in low eccentricity orbits.
This mechanism can reproduce 12 S-stars just by assuming that the binaries initially lie within the stellar disk as observed nowadays. To reproduce all the S-stars, the original disk must have been extended down to $0.006\pc$. However in this case many B- and O-type stars remain in low eccentricity orbits below $0.03\pc$, in contrast with the observations. Therefore, some other mechanism is necessary to disrupt the disk below $0.03\pc$.
This scenario can also explain the high eccentricity of the G-objects, if they have a stellar origin.
\end{abstract}%

\bibliographystyle{apj}





\section{Introduction}

The Galactic center harbours thousands of young stars within a parsec distance from Milky Way's supermassive black hole (SMBH), SgrA*.
Hundreds of Wolf-Rayet (WR) and O-type stars, with an estimated age of $6 \myr$, lie in the region between $0.03$ and $0.4\pc$ from SgrA*. A fraction of these young stars ($20$--$50\%$) appears to form a nearly Keplerian, eccentric ($ e  \simeq 0.3$) disk, named clockwise (CW) disk \citep{pau06,bar09,lu09,do13,yel14}.
{ The ensemble of the closest stars to the SMBH is called the S-star cluster. No WR and O-type stars have been observed among the S-stars, most of which are B-type stars \citep[32 out of 40,][]{gil17}. Out of the 32 B-type S-stars, 8 appear to be part of the CW disk, while the remaining 24 have randomly oriented orbits.}


The origin of the young stars is puzzling: their young age poses serious constraints on any dynamical migration scenario \citep[e.g.][]{por2003,kim2003,kim2004,fujii2008,fujii2009,fujii2010}; moreover, the tidal shear from the SMBH would disrupt molecular clouds, preventing \textit{in situ} star formation \citep[see][for a review]{map16b}.
However, it has been shown that star formation can still occur in a gaseous disk around the SMBH \citep{too64,nay05,nay07,col08}. The most accepted scenario for the formation of the CW disk is the infall and disruption of a molecular cloud, which settled into a gaseous disk around SgrA* \citep{bon08,map08,hob09,ali11,luc13,tra16,map16a}. This mechanism leads to the formation stars in mildly eccentric orbits, successfully reproducing the dynamical properties of the CW disk. However, this scenario fails to explain the highly eccentric, random orbits of the S-stars.



Many solutions have been suggested to explain the origin of the S-stars: binary breakup by the SMBH \citep{hil91,per09}, disk migration \citep{lev07}, Kozai-Lidov oscillations \citep{chen14,subr16} and fragmentation of active galactic nuclei outflow \citep{naya18}.

{
In the binary breakup scenario, the S-stars are captured by the SMBH via tidal disruption of binary stars. This mechanism can produce stars with very high eccentricities \citep[$e\sim 0.95$--$0.99$,][]{hills88}, which can then relax via scalar resonant relaxation towards a thermal eccentricity distribution, similar to the observed one \citep{per09,mad11,ant13b,ham14}}. Binaries can come either from the outer parsec or from a disk of stars between $0.04$ and $0.1\pc$. In the former case, binaries are scattered into radial orbits by a massive perturber \citep{per07,per08}, while in the latter case eccentricity is excited by Kozai-Lidov resonances, resulting in very small pericenter passage that allows the tidal disruption of the binary \citep{madi09,madi14,subr16}.  
However, these scenarios cannot explain the lack of WR/O-type stars in highly eccentric orbits.


Recently, \citet{hab17} analysed the combined spectroscopic data for 8 S-stars. They infer an age of $6.6\myr$ for the star S2 and less than $15\myr$ for the remaining S-stars. This is in tension with the binary break-up scenario, which requires the eccentricity to relax for at least $40\myr$ after the break-up of the binary \citep{baror18}.
Interestingly, the new age estimate for the S-stars is compatible with the age of the CW disk. While the S-stars and the CW disk appear to be nowadays two distinct populations, their similar age raises the question whether they were a single stellar population in the past. 

Finally, there is another class of highly eccentric objects orbiting around SgrA*: the G-objects \citep{gil12,gil13a,wit14,sha16,ple17}. These are faint, dusty objects visible in the L$^\prime$ band and Br$\gamma$ line, but lacking any K-band emission proper of a star. So far, only two objects, G1 and G2, have been observed, but more are expected to be found in the near-future. Several theories have been proposed to explain the nature of the G-objects, but only a few studies have tried to explain the origin of their high eccentricity \citep{mur12,tra16b}.

Here we investigate a new formation mechanism for the S-stars and the G objects, assuming that the S-stars and the CW disk were born in the same star formation episode, via the fragmentation of a gaseous disk.
There are at least 3 known binaries in the CW disk, and many more binary candidates exist \citep{pfu14,nao18}. 
It is also well known that a dark cusp of compact remnants is expected to have grown around the SMBH, via dynamical friction and \textit{in situ} star formation \citep{bah76,bah77,hop09,merr10,anto14,gen18,hail18}. 
{ In particular, \citet{alex09} predict that stellar black holes with mass ${\gtrsim}10 \msun$ will sink towards the SMBH and develop a steep cusp with a density power-law exponent of ${\sim}2$--$3$.}

Since binaries have a larger cross section, their encounter rate is enhanced with respect to single stars. It is therefore possible that an encounter can result in the ionization of the binary, kicking the ionized binary components into highly eccentric orbits. While the isolated three-body encounter has been studied in detail \citep[e.g.][]{heg75,hut83a,hut83b,hut93,hegg93,good93,hegg93,mcmill96,hegg96}, no studies were dedicated so far to the Keplerian three-body encounter, in which all encountering bodies lie in Keplerian orbits about a SMBH. 

In this paper, the first in the series, we investigate the formation of S-star via ionizing three-body encounters, assuming that both S-stars and the CW disk were born in the same star formation episode. 

In Section~\ref{sec:methods} we describe the numerical setup of our 4-body simulations. Section \ref{sec:results} presents our main results regarding the production of S-stars via ionizing encounters. In Section~\ref{sec:discussion}, we discuss the implications and \textit{caveats} of our work. Finally, our conclusions are summarized in Section~\ref{sec:conclusions}.

\section{Methods}\label{sec:methods}
We perform 4-body simulations in which a binary from the CW disk and a stellar black hole undergo a 3-body encounter.
We run four sets of realizations, referred as set A, B, A$_\mathrm{ex}$ and B$_\mathrm{ex}$. In set A and A$_\mathrm{ex}$, the binary components are WR/O-stars, while in set~B and B$_\mathrm{ex}$ are modelled as B-type stars.
For the sets \Aex and \Bex, we assume that the CW disk was more extended towards the SMBH in the past.

\subsection{Initial conditions}
The SMBH mass is set to $4.31\times10^6 \msun$ \citep{gil09a,gil17}.

The binary orbit about the SMBH is modeled following the observed properties of the CW disk \citep{bar09,do13,yel14}. The semimajor axis is drawn from a power-law distribution with index $-1.93$, in the range $0.03$--$0.1\pc$ for set A and B and in the range  $0.006$--$0.06\pc$ for set A$_\mathrm{ex}$ and B$_\mathrm{ex}$, consistent with the surface density $\Sigma(r) \propto -0.93$ reported by \citet{do13}. The eccentricity is drawn from a normal distribution with $\langle e\rangle = 0.3\pm0.1$. We fix the orbit of the binary in the $x$--$y$ plane and vary the orbital orientation of third body so that the encounter always occurs along the $x$ axis.

For the semimajor axis and eccentricity of the inner binary we adopt the distributions from \citet{san12}. The eccentricity distribution follows a power-law with index $-0.45$ between 0 and 1.
The period distribution follows $f(P) \propto (\log_{10}P)^{-0.55}$, with $\log_{10}P \in (0.15, 0.55)$ and $P$ is in days. We truncate the binary semimajor axis to the Hill radius at pericenter $$r_{\rm H} = 0.5\,a_{\rm bin} (1 - e_{\rm bin}) \left(\frac{m_{\rm bin}}{3M_{\rm SMBH}}\right)^{1/3}$$ if the semimajor axis exceeds $r_{\rm H}$. Likewise, we redraw the semimajor axis and eccentricity of the inner binary if they would immediately lead to a collision, i.e. $a_{\rm in} (1 - e_{\rm in}) < R_{1} + R_{2}$. All the others Keplerian elements $(i, \omega, \Omega, \nu)$ are randomly sampled. 

In set~A and A$_\mathrm{ex}$, the mass of the binary stars is randomly drawn from a power-law distribution with exponent $\alpha=-1.7$ between $25$ and $150 \msun$, consistent with the WR and O -type population of the CW disk \citep{lu13}.
In set~B and B$_\mathrm{ex}$, the mass of the binaries is uniformly sampled between $8$ and $14\msun$, representing the B-type population.
The stellar radius is set to $R = (M/{\rm M_\odot})^{0.8} \rsun$. 

Motivated by the LIGO detections and population synthesis studies \citep[e.g.][]{spera2018} we set the mass of the intercepting stellar black hole to $30 \msun$. Supplementary sets of simulations with different black holes masses ($m_\mathrm{bh} = 10$, $500$ and $1000 \msun$) can be found in the Appendix~\ref{sec:m10}. The orbital eccentricity of the stellar black hole about the SMBH is drawn from a thermal distribution. The orbital orientation is uniformly sampled over the sphere, and the azimuthal angle at encounter is randomly picked in the range allowed by the eccentricity. 
The impact parameter is a 3-dimensional vector drawn from sphere of radius $2\,a_{\rm in}$, where $a_{\rm in}$ is the semimajor axis of the binary, surrounding the center of mass of the binary.
We set the radius of the black hole to its tidal radius $R_{\rm roche} = \sqrt[3]{2\pi/3 m_{\rm bh}/\rho_\odot} \simeq 6.4 \rsun$, where $\rho_\odot = 1.41 \,\rm g/cm^3$ is the mean solar density. This lets us detect as collisions and exclude those simulations that would end in the tidal disruption of one of the stars. Table~\ref{tab:ic} lists the main initial conditions of our model.

We run $10^6$ realizations for each set. The simulations are run for about 1/8 the orbital period of the binary about the SMBH $T_{\rm bin}$, with the encounter occurring approximatively $T_{\rm bin}/16$ after the start of the simulations.

\begin{deluxetable*}{c|c|c|c|c}
	\tabletypesize{\scriptsize}
	\tablecaption{Initial setup of our simulations.\label{tab:ic}}
	\tablewidth{\linewidth}
	\tablehead{\colhead{Properties} & \colhead{Set A} & \colhead{Set B} & \colhead{Set A$_\mathrm{ex}$} & \colhead{Set B$_\mathrm{ex}$}}
		\startdata
		Realizations &  $10^6$ & $10^6$ & $10^6$ & $10^6$\\ 
		$a_{\rm bin}$ [pc] & \multicolumn{2}{c|}{$a^{-1.93}$, $a\in(0.03, 0.1)$} & \multicolumn{2}{c}{$a^{-1.93}$, $a\in(0.006, 0.06)$}\\
		$e_{\rm bin}$  & \multicolumn{2}{c|}{$\langle e\rangle = 0.3\pm0.1$} & \multicolumn{2}{c}{$\langle e\rangle = 0.3\pm0.1$} \\
		
		$a_{\rm in}$ & \multicolumn{2}{c|}{\citet{san12}} & \multicolumn{2}{c}{\citet{san12}} \\
		$e_{\rm in}$ & \multicolumn{2}{c|}{\citet{san12}} & \multicolumn{2}{c}{\citet{san12}}\\
		$m_{1}, m_{2}$ [$M_\mathrm{\odot}$] & $m^{-1.7}$, $m\in (25, 150)$ & $\text{unif}\in(8$--$14)\msun$ & $m^{-1.7}$, $m\in (25, 150)$ & $\text{unif}\in(8$--$14)$\\

		$e_{\rm sin}$ & \multicolumn{2}{c|}{$f(e) \propto e$} & \multicolumn{2}{c}{$f(e) \propto e$}\\
		$b$ & \multicolumn{2}{c|}{$2\,a_{\rm in}$} & \multicolumn{2}{c}{$2\,a_{\rm in}$}
		
		\enddata
		\tablecomments{\footnotesize Row~1: semimajor axis of the binary orbit about the SMBH; 
			row~2: eccentricity of the binary orbit about the SMBH; 
			row~3: semimajor axis of the inner binary; 
			row~4: eccentricity of the inner binary; 
			row~5: mass of the binary components; 
			row~6: semimajor axis of the single body orbit about the SMBH; 
			row~7: eccentricity of the single body orbit about the SMBH; 
			row~8: impact parameter of the single star about the binary center of mass.
		}
\end{deluxetable*}

\subsection{Setting up a Keplerian two-body encounter}

Setting up a three-body encounter in a Keplerian potential is not as straightforward as in the isolated case. Here we describe how we set up an encounter between two bodies in Keplerian orbits, and how we map eccentricity and semimajor axis about the central SMBH to velocity and impact parameter at the encounter.

Consider two encountering bodies $\cal A$ and $\cal B$. 
We first fix the encounter position in space $\mathbf{R}_{\rm A}$, which has to lie along the orbit of body $\cal A$. Then we choose a velocity vector $\mathbf{V}_{\cal A}$ at encounter position $\mathbf{R}_{\rm A}$ that is consistent with semimajor axis, eccentricity and orbital orientation of $\cal A$. From $\mathbf{V}_{\cal A}$ and $\mathbf{R}_{\rm A}$ we can then compute the full set of Keplerian orbital parameters of $\cal A$ $(a, e, i, \omega, \Omega, \nu)_{\cal A}$. We repeat the same steps for body $\cal B$ using $\mathbf{R}_{\cal B} = \mathbf{R}_{\rm A} + \mathbf{B}$, where $\mathbf{B}$ is a chosen impact parameter vector. Once a consistent velocity vector $\mathbf{V}_{\cal B}$ is also chosen for $\cal B$, we can compute its six Keplerian orbital parameters $(a, e, i, \omega, \Omega, \nu)_{\cal B}$. 

We then shift the true anomaly $\nu$ back in time by solving the Kepler equation twice for each body. In this way, we ensure that an encounter will occur between the two bodies.
Afterwards, we can convert the new Keplerian elements $(a, e, i, \omega, \Omega, \nu')$ of each body to Cartesian coordinates for the numerical integrator. 

\subsection{Mikkola's Algorithmic Regularization code}
We run the simulations using \texttt{TSUNAMI}, an implementation of Mikkola's algorithmic regularization \citep[MAR,][]{mik99a,mik99b}. This code is particularly suitable for studying the dynamical evolution of few-body systems in which strong gravitational encounters are very frequent and the mass ratio between the interacting objects is large. The MAR scheme solves the equation of motions derived from a time-transformed Hamiltonian, for which the timestep does not go to $0$ for $r\rightarrow{}0$ (see \citealt{mik99a} for the details).

Since the timestep evaluations are sparse in physical time, the MAR scheme can potentially allow for particle interpenetration even when checking for collisions at each timestep. Therefore, we implemented a collision checking algorithm that uses the predicted pericenter passage during close encounters.

\texttt{TSUNAMI} uses a second order leapfrog scheme in combination with the Bulirsh-Stoer extrapolation algorithm \citep{sto80} to increase the accuracy of the numerical results. Our code includes velocity-dependent forces following the algorithm described in \citet{mik06,mik08}. Among these, we included the post-Newtonian terms 1PN, 2PN and 2.5PN \citep{blanchet2006} and the tidal drag-force from \citet{sam18a}, although these terms are not switched on in the present work.

\texttt{TSUNAMI} integrates the equations of motion employing relative coordinates by means of the so called chain structure. This change of coordinates reduces round-off errors significantly \citep{aar03}. 

More details on the \texttt{TSUNAMI} code will be presented in a following work (Trani A.A. et. al, in preparation).

\section{Results}\label{sec:results}
\begin{figure}[h]
	\includegraphics[width=\linewidth]{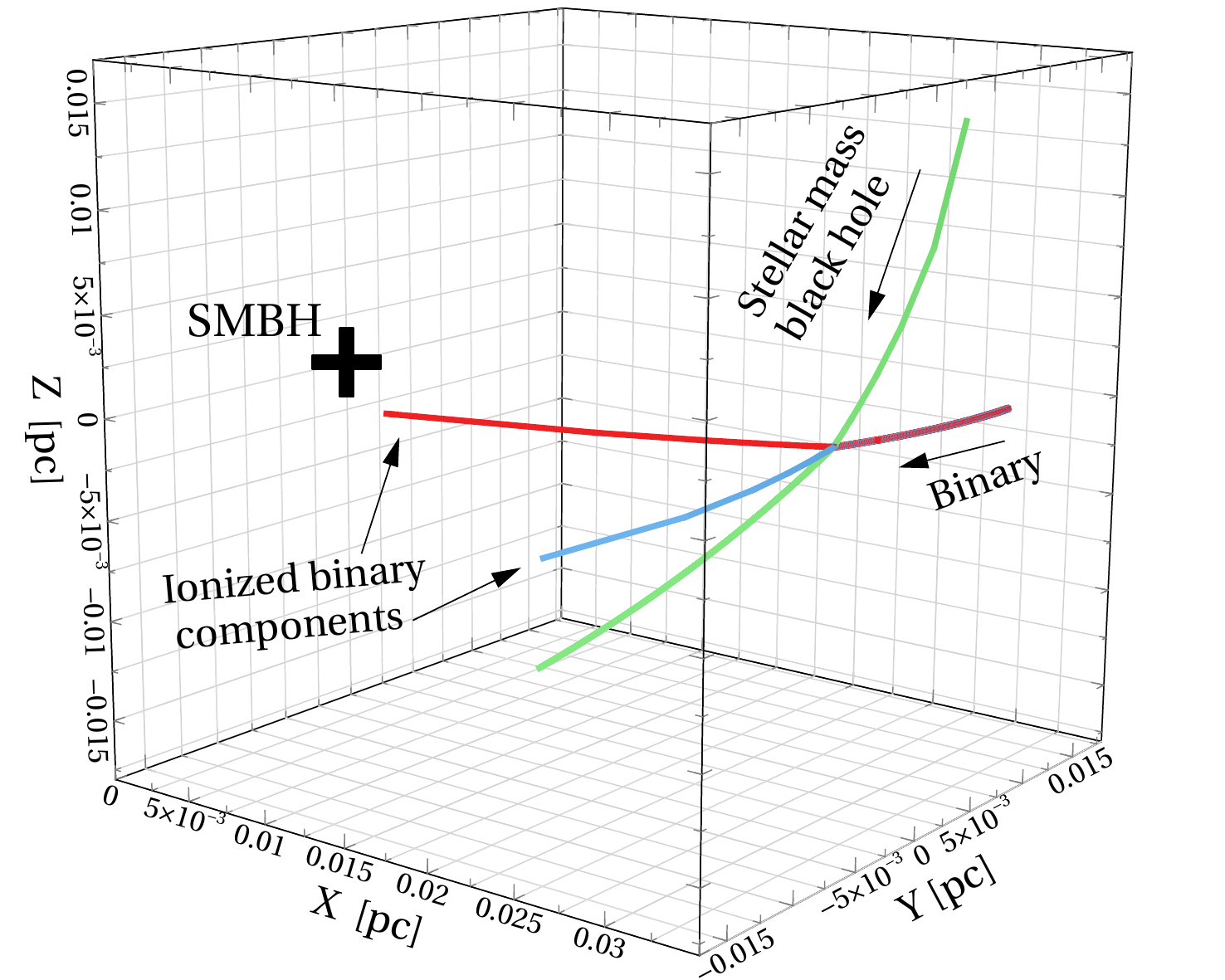}
	\caption{Trajectories of the bodies in individual realization, in the reference frame at rest with the SMBH. Black cross: SMBH. Green line: stellar black hole. Blue and red lines: binary star members.}
	\label{fig:single}
\end{figure}

Figure~\ref{fig:single} shows the trajectory of the bodies in a single realization. In this particular realization, the binary is ionized and the binary components are scattered into eccentric orbits. 

Ionization occurs in $4.70\%$ and $24.50\%$ of the runs of set~A~and~B, respectively. If the binary remains bound, we classify this outcome as flyby. Other kind of outcomes may occur: collisions/tidal disruptions between stars and the stellar black hole; exchanges, in which a binary member is exchanged with the black hole; and ejections, in which any of the bodies is ejected from system and becomes unbound with respect to the SMBH.
Table~\ref{tab:result} summarizes the outcomes of the simulations.

\begin{figure}[htb]
	\includegraphics[width=\linewidth]{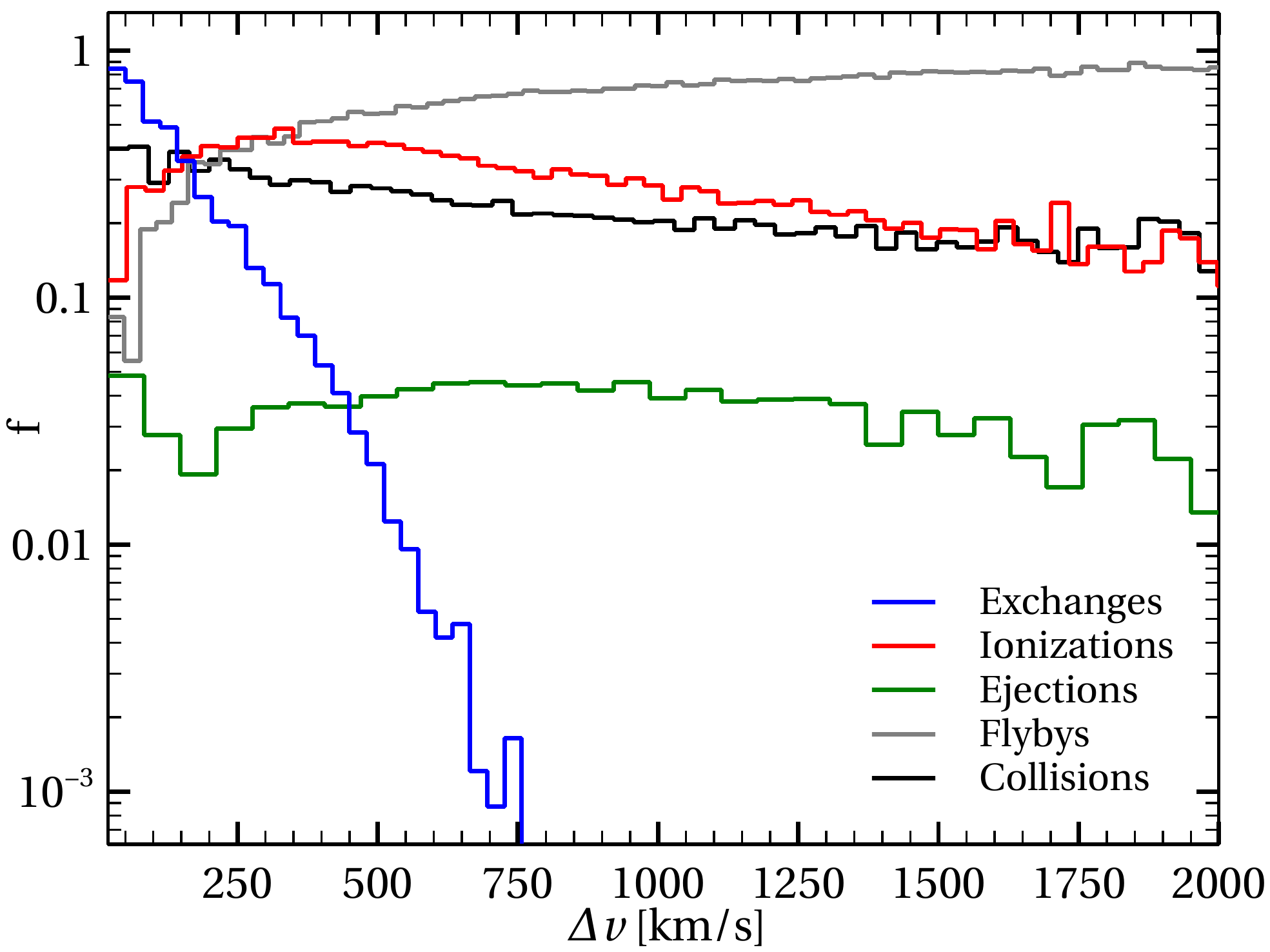}
	\caption{Outcome fraction as a function of the relative velocity at encounter between the binary and the stellar black hole, for set~B. Red line: ionizations. Grey line: flybys. Blue line: exchanges. Green line: ejections.
	}
	\label{fig:outmap}
\end{figure}

The outcome crucially depends on the relative velocity at encounter between the binary centre of mass and the stellar black hole, as shown in Figure~\ref{fig:outmap}. For low relative velocity, most encounters lead to exchanges. As the relative velocity increases, the total energy of the three-body system becomes positive and ionizations become possible. However, for higher relative velocity the encounter is very rapid and little energy is exchanged, so that the ionization cross section rapidly falls off \citep{hut83a,hut83b}.

Note that the relative velocity is mainly due to the relative orbital orientation of the binary and the black hole, with some additional velocity dispersion given by the orbital eccentricity. Low relative velocity results when the encountering bodies orbit in the same plane and direction, while high velocity dispersion results from head-on encounters. The peak of ionizations occurs when binary and stellar black hole are mutually inclined by ${\approx}21^\circ$.

\begin{figure}[htb]
	\includegraphics[width=\linewidth]{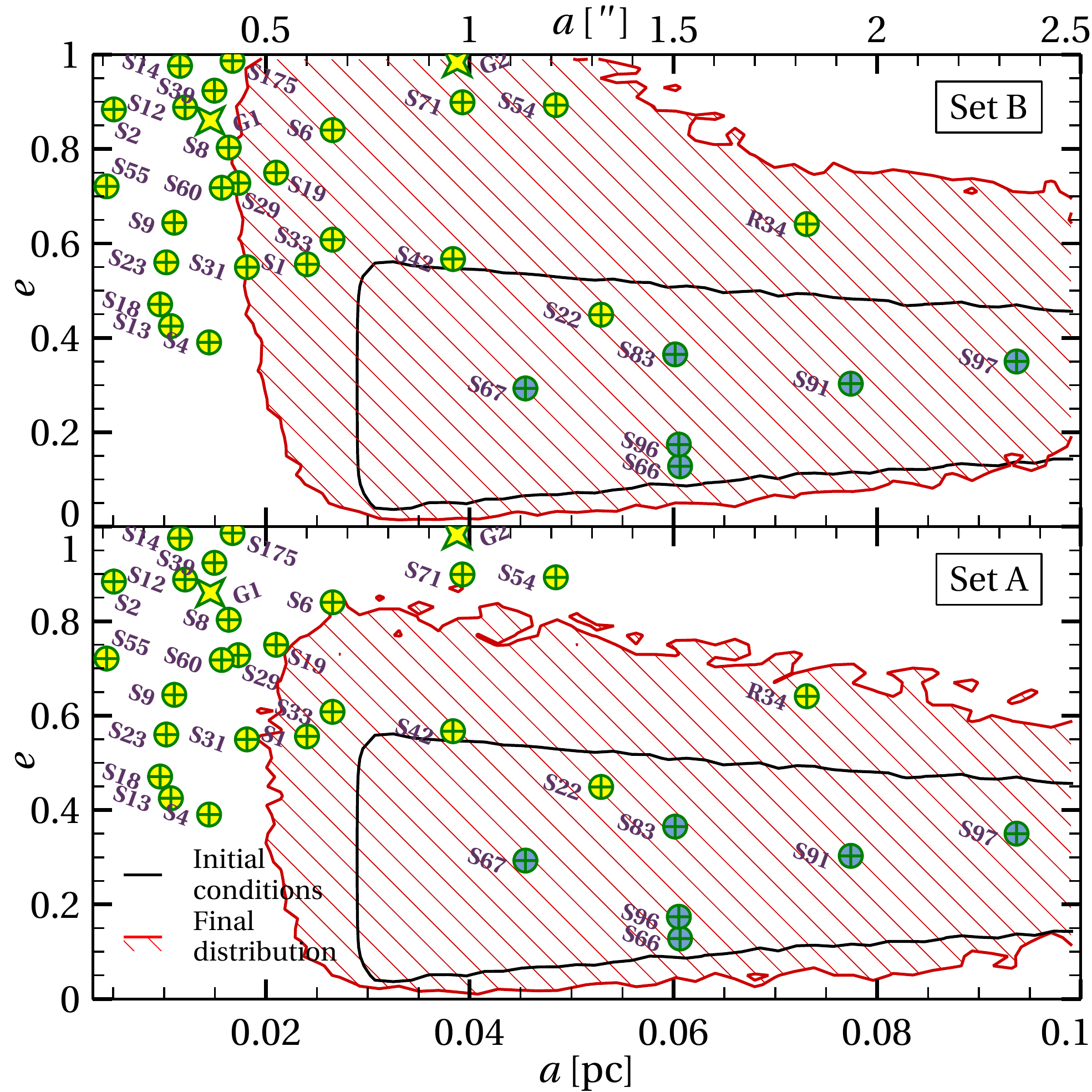}
	\caption{ 
		Semimajor axis-eccentricity map of ionized binary components in set~B (top) and A (bottom). Blue circles with green cross: S-stars classified by \citet{gil17} as part of the CW disk.
		Yellow circles with green cross: early type S-stars not classified as part of the CW disk. Yellow stars with green contour: G1 and G2 objects. Black contour: initial conditions. Red filled contour: $2\sigma$-cut final distribution.
	}
	\label{fig:aemap}
\end{figure}

In Figure~\ref{fig:aemap} we show the semimajor axis and eccentricity of the ionized binary components for set~A and B, along with the observed parameters of the B-type S-stars, and the known G-objects.

\begin{deluxetable}{lcccc}
	\tabletypesize{\scriptsize}
	\tablecaption{Outcome of the simulations.\label{tab:result}}
	\tablewidth{\linewidth}
	\tablehead{\colhead{Outcomes} & Set~A & Set~B & Set~\Aex & Set~\Bex }
	\startdata
 	Ionization & $4.70\%$ & $24.50\%$ & $2.71\%$  & $18.95\%$ \\
 	Collision & $17.14\%$ & $23.14\%$ & $14.84\%$ & $20.15\%$ \\
 	Flyby & $77.50\%$ & $50.77\%$  & $82.27\%$ & $60.51\%$ \\
 	Exchange & $0.71\%$ & $1.58\%$ & $0.21\%$ & $0.40\%$ \\
 	Ejection & $4.11\%$ & $3.03\%$ & $1.86\%$ & $1.04\%$
	\enddata
\end{deluxetable}

In set~A, the ionized stars remain at mild eccentricity, comparable to the that of the original binaries. O and Wolf-Rayet stars are too massive to receive a strong kick from the stellar black hole. Therefore, the O/WR stars from set~A can match the orbital properties of only a few of the low-eccentricity S-stars, and none of the G-objects. 

In contrast, the B-stars from set~B get scattered into higher eccentricity orbits.

While the initial eccentricity of the binaries does not exceed $0.6$, the ionized binary components distribution has a tail with $0.6$--$0.99$ eccentricity. In particular, the semimajor axis and eccentricity of ionized stars from set~B is compatible with the orbits of 12 S-stars in the semimajor axis range $0.016$--$0.075\pc$ (S1, S6, S8, S19, S29, S31, S33, S42, S54, S60, S71, R34)\footnote{ Note that S22 is already compatible with the initial conditions, therefore we do not include it in our list. The reason why \citet{gil17} do not classify it as part of the CW disk is because S22 has a different orbital orientation with respect to the CW disk. While our model is not able to reproduce tilting/disruption of the CW disk, several authors pointed out the mechanisms to explain the presence of young stars outside the disk \citep[e.g.][]{sub09,ali11,luc13,tra16}.
} and the G2 object. The remaining S-stars have an exceedingly small semimajor axis compared to the ionized binaries of set~B.

\begin{figure}[htb]
	\includegraphics[width=\linewidth]{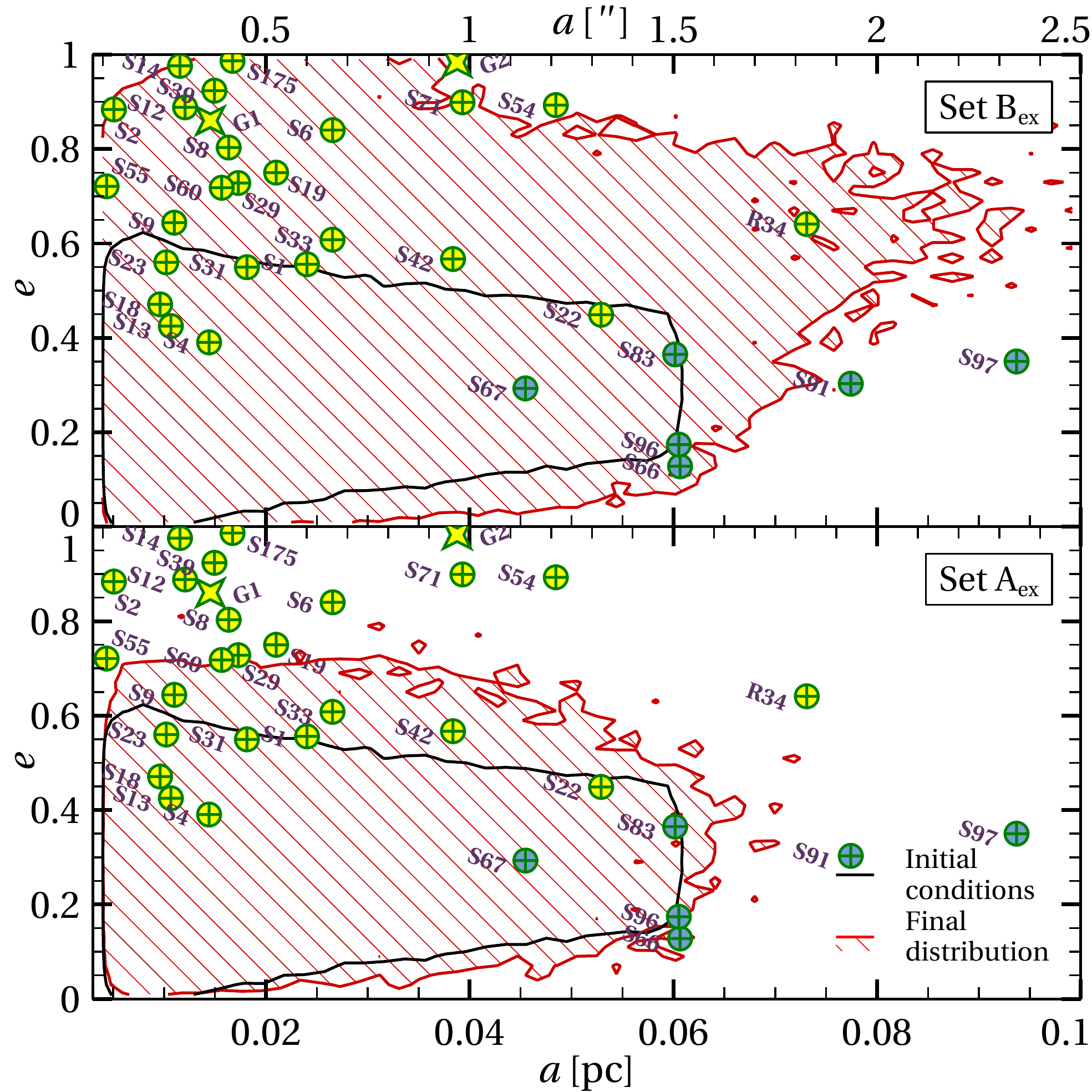}
	\caption{
		Same as Figure~\ref{fig:aemap}, but for set~B$_\mathrm{ex}$ (top) and A$_\mathrm{ex}$ (bottom).
	}
	\label{fig:aemap_ex}
\end{figure}

Figure~\ref{fig:aemap_ex} shows the semimajor axis-eccentricity map for the ionized stars of set A$_\mathrm{ex}$ and B$_\mathrm{ex}$, in which the CW disk is extended down to $0.006\pc$.
The distribution of ionized binary components from set~\Bex overlaps with that of the S-stars. As in set~A, stars from set \Aex cannot reproduce S-stars with eccentricity greater than ${\sim}0.7$.

\section{Discussion}\label{sec:discussion}

No star from set~B can match the semimajor axis of the S-stars with semimajor axis smaller than $0.016\pc$. This is not surprising, since the orbits before and after the scattering event have to be crossing, i.e. the apocenter of the final orbit must be larger than the pericenter of the initial orbit. This sets a constrain on the initial binary orbit that depends on the final apocenter distance. The star S55 has an apocenter of $0.007\pc$, the smallest apocenter among the S-stars. In order to reproduce its orbit via scattering, the initial orbit must have a semimajor axis of  at least $0.01\pc$, assuming an initial eccentricity of $0.3$.
Therefore, it is not possible to reproduce the innermost S-stars via this mechanism without assuming that the disk was more extended in the past. 

One issue with our scenario is the abundance of stars that remain at lower eccentricity: only a few percent of stars end up in highly eccentric orbits. In other words, most stars remain within the initial disk. This is not an issue in the case of set~A and B, since the distribution of surviving binaries follows the CW disk properties as observed nowadays.

\begin{figure}[htb]
	\includegraphics[width=\linewidth]{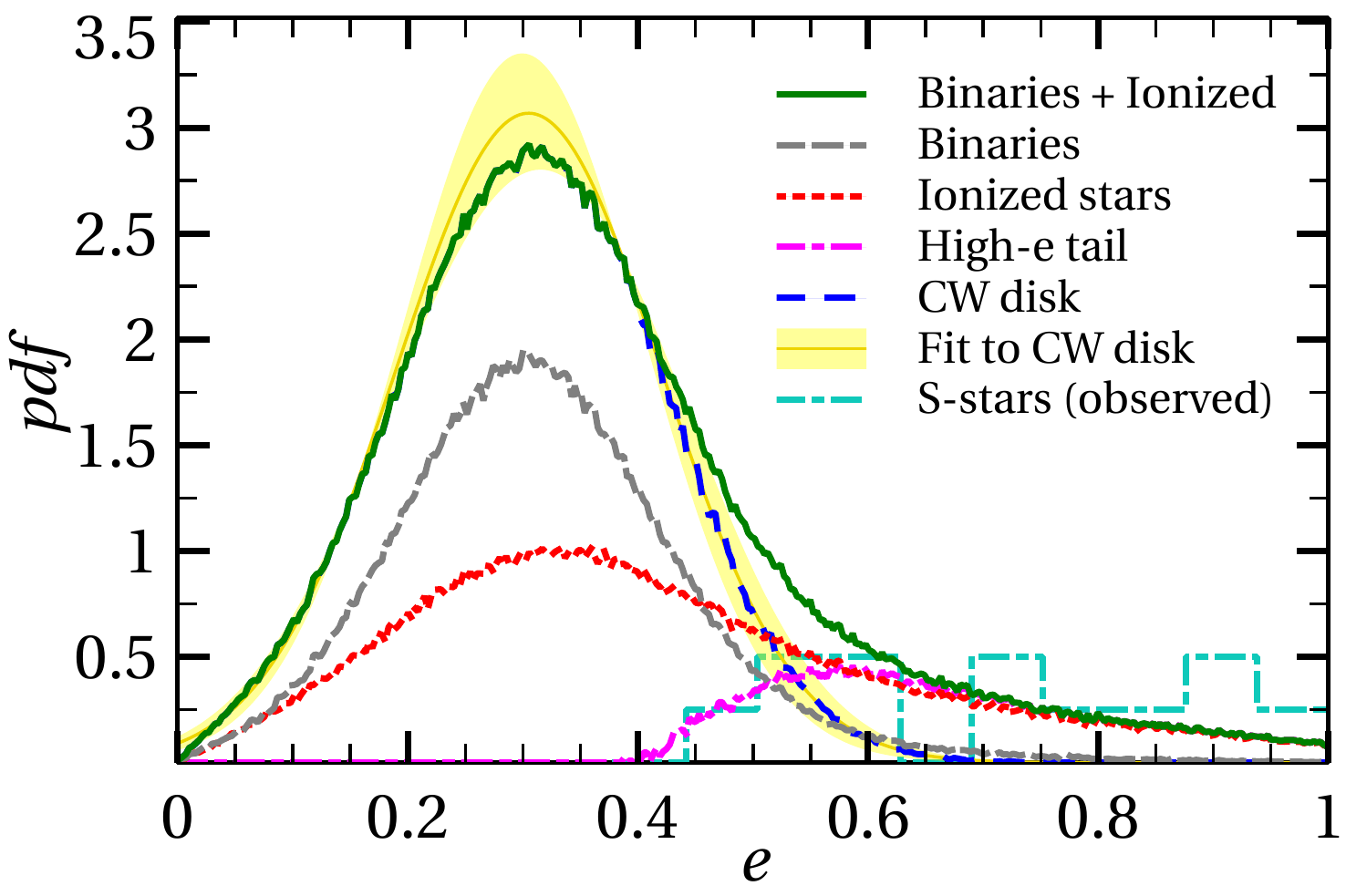}
	\caption{ 
		Eccentricity distributions of stars in the semimajor axis range $0.016$--$0.075\pc$. Green solid line: all ionized stars and survived binaries from set~B. Grey dashed line: survived binary stars. Red dotted line: ionized binary stars. Yellow line with boostrapped 1$\sigma$ confidence band: normal fit to all stars and binaries with eccentricity less than $0.5$. Blue short-dashed line: ionized and binaries stars consistent with the normal fit (consistent with CW disk distribution).
		Magenta dot-dashed line: ionized and binary stars not consistent with the normal fit. Cyan dot-dashed line: observed distribution of the S-stars in the considered semimajor axis range (arbitrary scale). All the other distributions are normalized so that the total distribution of binaries and ionized stars (green solid line) is normalized to one.
	}
	\label{fig:edist}
\end{figure}

	This is clear from Figure~\ref{fig:edist}, which compares the eccentricity distributions of stars from set~B with that of the S-stars with semimajor axis between $0.016$ and $0.075\pc$. In this range, $46\%$ of the stars are survived binary stars while ionized binary components constitute the remaining $54\%$. This is slightly more than $2\times 24.50 = 49\%$ from the total results reported in Table\ref{tab:result}).
	
	The bulk of the stars remains at mild eccentricity, consistent with the observed distribution of CW disk stars. However, the high-eccentricity tail is not consistent with the CW disk. 
	
	In order to isolate the high-tail component, we fit the eccentricity of all ionized and binary stars with $e < 0.5$ to a normal distribution, obtaining a value consistent to the initial one ($\mu = 0.2993$, $\sigma = 0.1053$). We then isolate the subset of data consistent with the obtained normal distribution by Monte Carlo sampling.
	We find that for set~B the high eccentricity tail consists $15\%$ of the total stars, of which $14\%$ are single stars and $1\%$ are binaries.
	Thus, an average of ${\approx}86$ B-stars must have resided in binaries to produce the 12 S-stars observed in the region.
	
	Current spectroscopic studies are limited down to ${\sim}10 \msun$ \citep{do13}. As such, the number of B-type stars in the region is largely unconstrained. \citet{lu13} estimates a total cluster mass between $1.4\times 10^4$ and $3.7\times 10^4 \msun$ assuming an initial mass function slope of $\alpha = -1.7$, extrapolated down to $1\msun$. Their estimate gives an average of $194$ B-type stars, which is more than a factor of $2$ larger than our estimate above, assuming $100\%$ binary fraction. Therefore, our scenario is consistent with the current state of observations. A better constrain on the B-type population in the Galactic center will be possible in the near-future with the advent of 30m-class telescopes \citep[i.e. the Extremely Large Telescope, ELT, and the Thirty Meter Telescope, TMT; see][]{gul14,do14}. 	
	

However, it is an issue for set~\Aex and \Bex: the presently observed disk extends only down to $0.03\pc$, not $0.006\pc$. Several authors have pointed out various processes that can affect the evolution of the disk: Kozai-Lidov resonance with a secondary disk \citep{loc09,loc09b} or with a gas torus \citep{sub09,haa11a,haa11b,tra16}, two-body relaxation \citep{sub14}, Kozai-Lidov mechanism within the disk itself \citep{chen14,subr16} and resonant relaxation \citep{rau96,koc15,baror18}. 
However, none of these mechanism is able to disrupt the disk below a certain radius. 

The simulations show that Keplerian three-body encounters can strongly affect the semimajor axis and eccentricity distribution of the stars about the SMBH. On the other hand, they can barely affect the orientation of the ionized binary components. Scattered stars mostly inherit the orbital orientation of the parent binary. However, this is not an issue for our scenario. Vector resonant relaxation can in fact randomize the orbits of the S-stars below $0.03\pc$ in a few million years \citep[][see also \citealt{baror18}]{hop06}. 

Note that not all the young stars in the outer $0.030\pc$ of the Galactic center are observed within the disk. Current estimate of the disk fraction varies between 20 and 50\% \citep{bar09,yel14}. Our numerical setup does not include any torque due additional components (e.g. an outer gaseus torus) able to disrupt or tilt the disk. Therefore, we do not model the change of orientation of the young binaries and stars from the original stellar disk. 

As shown in Figure~\ref{fig:outmap} and Table~\ref{tab:result}, a large fraction of binaries survive the three-body encounter. These binaries have their orbital properties altered by the three-body encounter. This can have a strong impact on the production of binary mergers, including the triggering of gravitational-waves-induced coalescence for binaries of compact remnants. More details on this topic will be presented in the next paper of this series.


Recently, \citet{szo18} found that vector resonant relaxation in galactic nuclei tends to redistribute massive remnants in a disk configuration. If their result applies to the Galactic center, it would strongly affect the outcome type of the three-body encounters, depending on the relative inclination between the stellar and the remnant disk. If the disks were aligned, the low velocity dispersion would decrease the chance of ionization and increase exchanges and tidal disruptions (Figure~\ref{fig:outmap}). In contrast, a counter-aligned disk would result in increased flybys. A disk misalignment of $20^\circ$--$30^\circ$ would maximize the number of ionizing encounters. Note that a misaligned disk would likely induce Kozai-Lidov oscillation in the disk, albeit damped by the spherical cusp of old stars.

It is worth noting that the eccentricity of the S-stars increases towards small semimajor axis. This is agreement with the predictions of our model: since the encounter rate increases towards the center, it is natural to expect higher chance of producing stars in highly highly eccentric orbits.

\begin{figure}[t]
	\includegraphics[width=\linewidth]{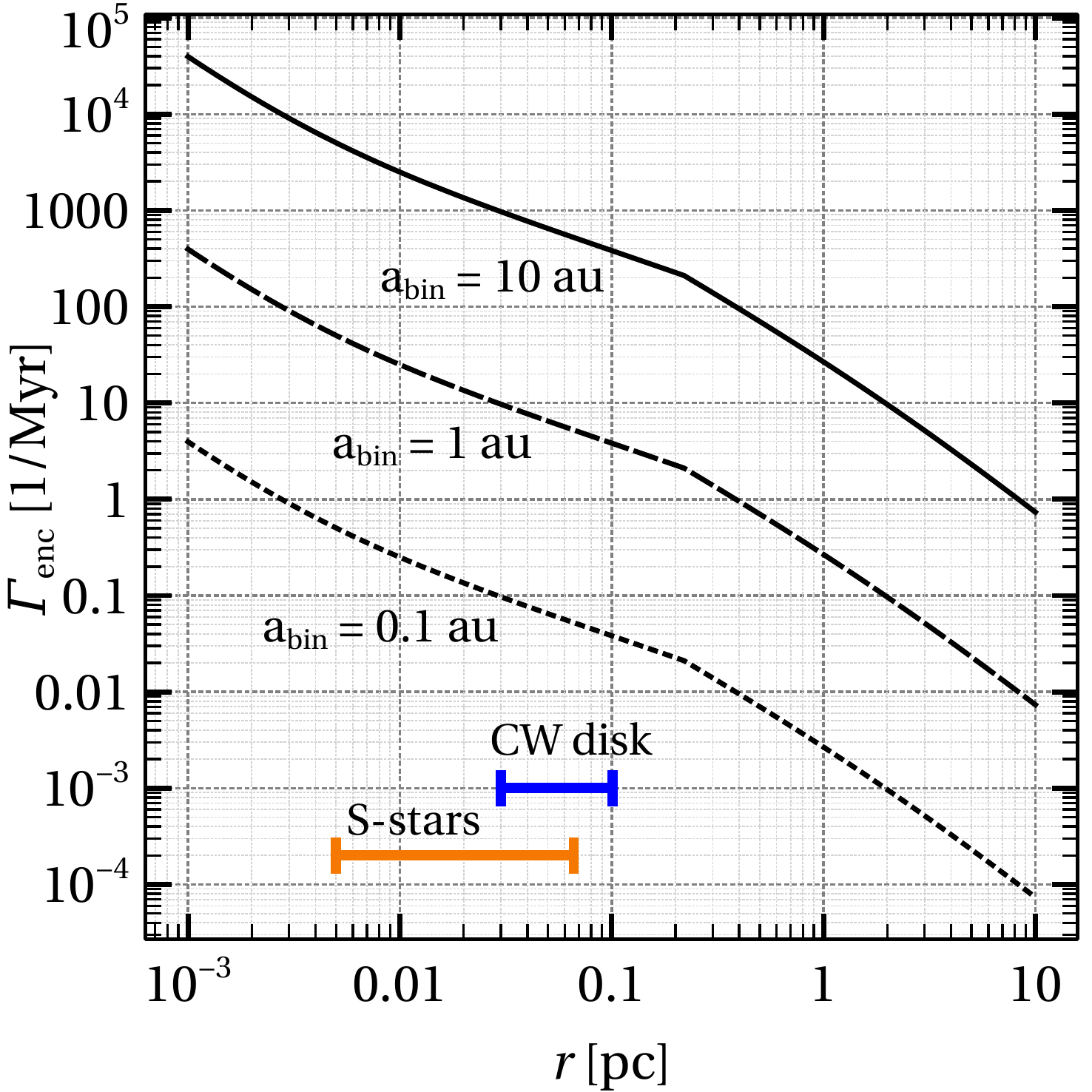}
	\caption{Encounter rate as a function of distance from the SMBH for a binary semimajor axis of $0.1$ (dotted line), $1$ (dashed line) and $10 \au$ (solid line). Using stellar density estimate from \citet{sch07}. The blue line indicates the present location of the CW disk. The orange line highlights the region of the observed S-stars.
	}
	\label{fig:rate}
\end{figure}
\subsection{Rates and timescales}\label{sec:rates}

In Figure~\ref{fig:rate} we show the encounter rate between binaries and single stars as a function of radius from the SMBH, computed using equation~11 from \citet{lei16a}. For the density profile, we use the broken power-law of \citet{sch07}. Note that both stellar and compact remnant distribution below $0.1\pc$ is highly uncertain, so also the derived encounter rates suffer from the same uncertainties. Nonetheless, the Galactic center is expected to host ${\sim}10000$ black holes in its central parsec from both theoretical considerations and observational evidence \citep{bah76,bah77,merr10,hail18}.

{
A binary formed in the Galactic center can undergo $1$--$10^3$ encounters in less than $6$--$15\myr$, which is the recent estimated age of the S-stars \citep{hab17}. Therefore, the binaries that survive the first encounter can be ionized during subsequent encounters. In the Appendix~\ref{sec:repeat} we present supplementary sets of simulations that follow the surviving binaries from sets~A~and~B undergoing repeated encounters.}

\section{Conclusions}\label{sec:conclusions}
We have run 4-body simulations of three-body encounters between binary stars and stellar black holes orbiting the SMBH in our Galactic center. We assume that both the S-stars and the CW disk stars were born in binaries in the same nearly Keplerian disk around SgrA*. 
We consider binaries composed of O/WR type or B-type stars, undergoing an encounter with $30\msun$ black holes.

B-type binaries can be easily ionized by the encounter, and their components can get scattered into highly eccentric orbits. On the other hand, O/WR type binaries are less easily disrupted, and the ionized stars remain in low eccentricity orbits.
	
{
	We can reproduce the orbits of 12 S-stars and the G2 object just by assuming that the initial binaries lie in the CW disk as observed nowadays. To reproduce the S-stars below $0.016\pc$, we need to extend the initial binary distribution down to $0.006\pc$. Even though in this way we can reproduce the whole population of S-stars, the simulations also predict a low-eccentricity population of B- and O-type stars within the inner $0.5$ arcsec of the Galactic center, in contrast with current observations.
}

These findings would suggest that the population of S-stars below $0.016\pc$ is a different population from the stars of the CW disk, despite their similar age. A single origin for both the S-stars and the CW disk via this scenario can be plausible if a mechanism to disrupt the stars in the disk below $0.03\pc$ is provided.


\acknowledgements
We thank the referee for insightful suggestions which improved the manuscript.
It is a pleasure to thank Tjarda Boekholt, Rainer Sch\"odel and Andrea Ghez for helpful discussions. This work was supported by JSPS KAKENHI Grant Number 17F17764. The initial conditions were generated using the AMUSE framework \citep{por09,por13,pel13}. All plots were made with the Veusz plotting package. The simulations were run on the calculation server at the Center for Computational Astrophysics at NAOJ.

\bibliography{ms.bib}

\begin{thebibliography}{}
\expandafter\ifx\csname natexlab\endcsname\relax\def\natexlab#1{#1}\fi

\bibitem[{{Aarseth}(2003)}]{aar03}
{Aarseth}, S.~J. 2003, {Gravitational N-Body Simulations}, 430

\bibitem[{{Alexander} \& {Hopman}(2009)}]{alex09}
{Alexander}, T., \& {Hopman}, C. 2009, \apj, 697, 1861

\bibitem[{{Alig} {et~al.}(2011){Alig}, {Burkert}, {Johansson}, \&
  {Schartmann}}]{ali11}
{Alig}, C., {Burkert}, A., {Johansson}, P.~H., \& {Schartmann}, M. 2011,
  \mnras, 412, 469

\bibitem[{{Antonini}(2014)}]{anto14}
{Antonini}, F. 2014, \apj, 794, 106

\bibitem[{{Antonini} \& {Merritt}(2013)}]{ant13b}
{Antonini}, F., \& {Merritt}, D. 2013, \apjl, 763, L10

\bibitem[{{Bahcall} \& {Wolf}(1976)}]{bah76}
{Bahcall}, J.~N., \& {Wolf}, R.~A. 1976, \apj, 209, 214

\bibitem[{{Bahcall} \& {Wolf}(1977)}]{bah77}
---. 1977, \apj, 216, 883

\bibitem[{{Bar-Or} \& {Fouvry}(2018)}]{baror18}
{Bar-Or}, B., \& {Fouvry}, J.-B. 2018, \apjl, 860, L23

\bibitem[{{Bartko} {et~al.}(2009){Bartko}, {Martins}, {Fritz}, {Genzel},
  {Levin}, {Perets}, {Paumard}, {Nayakshin}, {Gerhard}, {Alexander},
  {Dodds-Eden}, {Eisenhauer}, {Gillessen}, {Mascetti}, {Ott}, {Perrin},
  {Pfuhl}, {Reid}, {Rouan}, {Sternberg}, \& {Trippe}}]{bar09}
{Bartko}, H., {Martins}, F., {Fritz}, T.~K., {et~al.} 2009, \apj, 697, 1741

\bibitem[{{Blanchet}(2006)}]{blanchet2006}
{Blanchet}, L. 2006, Living Reviews in Relativity, 9, 4

\bibitem[{{Bonnell} \& {Rice}(2008)}]{bon08}
{Bonnell}, I.~A., \& {Rice}, W.~K.~M. 2008, Science, 321, 1060

\bibitem[{{Chen} \& {Amaro-Seoane}(2014)}]{chen14}
{Chen}, X., \& {Amaro-Seoane}, P. 2014, \apjl, 786, L14

\bibitem[{{Chu} {et~al.}(2018){Chu}, {Do}, {Hees}, {Ghez}, {Naoz}, {Witzel},
  {Sakai}, {Chappell}, {Gautam}, {Lu}, \& {Matthews}}]{chu18}
{Chu}, D.~S., {Do}, T., {Hees}, A., {et~al.} 2018, \apj, 854, 12

\bibitem[{{Collin} \& {Zahn}(2008)}]{col08}
{Collin}, S., \& {Zahn}, J.-P. 2008, \aap, 477, 419

\bibitem[{{Do} {et~al.}(2013){Do}, {Lu}, {Ghez}, {Morris}, {Yelda}, {Martinez},
  {Wright}, \& {Matthews}}]{do13}
{Do}, T., {Lu}, J.~R., {Ghez}, A.~M., {et~al.} 2013, \apj, 764, 154

\bibitem[{{Do} {et~al.}(2014){Do}, {Wright}, {Barth}, {Barton}, {Simard},
  {Larkin}, {Moore}, {Wang}, \& {Ellerbroek}}]{do14}
{Do}, T., {Wright}, S.~A., {Barth}, A.~J., {et~al.} 2014, \aj, 147, 93

\bibitem[{{Fujii} {et~al.}(2008){Fujii}, {Iwasawa}, {Funato}, \&
  {Makino}}]{fujii2008}
{Fujii}, M., {Iwasawa}, M., {Funato}, Y., \& {Makino}, J. 2008, \apj, 686, 1082

\bibitem[{{Fujii} {et~al.}(2009){Fujii}, {Iwasawa}, {Funato}, \&
  {Makino}}]{fujii2009}
---. 2009, \apj, 695, 1421

\bibitem[{{Fujii} {et~al.}(2010){Fujii}, {Iwasawa}, {Funato}, \&
  {Makino}}]{fujii2010}
---. 2010, \apjl, 716, L80

\bibitem[{{Generozov} {et~al.}(2018){Generozov}, {Stone}, {Metzger}, \&
  {Ostriker}}]{gen18}
{Generozov}, A., {Stone}, N.~C., {Metzger}, B.~D., \& {Ostriker}, J.~P. 2018,
  \mnras, 478, 4030

\bibitem[{{Gillessen} {et~al.}(2009){Gillessen}, {Eisenhauer}, {Trippe},
  {Alexander}, {Genzel}, {Martins}, \& {Ott}}]{gil09a}
{Gillessen}, S., {Eisenhauer}, F., {Trippe}, S., {et~al.} 2009, \apj, 692, 1075

\bibitem[{{Gillessen} {et~al.}(2012){Gillessen}, {Genzel}, {Fritz}, {Quataert},
  {Alig}, {Burkert}, {Cuadra}, {Eisenhauer}, {Pfuhl}, {Dodds-Eden}, {Gammie},
  \& {Ott}}]{gil12}
{Gillessen}, S., {Genzel}, R., {Fritz}, T.~K., {et~al.} 2012, \nat, 481, 51

\bibitem[{{Gillessen} {et~al.}(2013){Gillessen}, {Genzel}, {Fritz},
  {Eisenhauer}, {Pfuhl}, {Ott}, {Cuadra}, {Schartmann}, \& {Burkert}}]{gil13a}
---. 2013, \apj, 763, 78

\bibitem[{{Gillessen} {et~al.}(2017){Gillessen}, {Plewa}, {Eisenhauer}, {Sari},
  {Waisberg}, {Habibi}, {Pfuhl}, {George}, {Dexter}, {von Fellenberg}, {Ott},
  \& {Genzel}}]{gil17}
{Gillessen}, S., {Plewa}, P.~M., {Eisenhauer}, F., {et~al.} 2017, \apj, 837, 30

\bibitem[{{Goodman} \& {Hut}(1993)}]{good93}
{Goodman}, J., \& {Hut}, P. 1993, \apj, 403, 271

\bibitem[{{Gullieuszik} {et~al.}(2014){Gullieuszik}, {Greggio}, {Falomo},
  {Schreiber}, \& {Uslenghi}}]{gul14}
{Gullieuszik}, M., {Greggio}, L., {Falomo}, R., {Schreiber}, L., \& {Uslenghi},
  M. 2014, \aap, 568, A89

\bibitem[{{Haas} {et~al.}(2011{\natexlab{a}}){Haas}, {{\v S}ubr}, \&
  {Kroupa}}]{haa11a}
{Haas}, J., {{\v S}ubr}, L., \& {Kroupa}, P. 2011{\natexlab{a}}, \mnras, 412,
  1905

\bibitem[{{Haas} {et~al.}(2011{\natexlab{b}}){Haas}, {{\v S}ubr}, \&
  {Vokrouhlick{\'y}}}]{haa11b}
{Haas}, J., {{\v S}ubr}, L., \& {Vokrouhlick{\'y}}, D. 2011{\natexlab{b}},
  \mnras, 416, 1023

\bibitem[{{Habibi} {et~al.}(2017){Habibi}, {Gillessen}, {Martins},
  {Eisenhauer}, {Plewa}, {Pfuhl}, {George}, {Dexter}, {Waisberg}, {Ott}, {von
  Fellenberg}, {Baub{\"o}ck}, {Jimenez-Rosales}, \& {Genzel}}]{hab17}
{Habibi}, M., {Gillessen}, S., {Martins}, F., {et~al.} 2017, \apj, 847, 120

\bibitem[{{Hailey} {et~al.}(2018){Hailey}, {Mori}, {Bauer}, {Berkowitz},
  {Hong}, \& {Hord}}]{hail18}
{Hailey}, C.~J., {Mori}, K., {Bauer}, F.~E., {et~al.} 2018, \nat, 556, 70

\bibitem[{{Hamers} {et~al.}(2014){Hamers}, {Portegies Zwart}, \&
  {Merritt}}]{ham14}
{Hamers}, A.~S., {Portegies Zwart}, S.~F., \& {Merritt}, D. 2014, \mnras, 443,
  355

\bibitem[{{Heggie}(1975)}]{heg75}
{Heggie}, D.~C. 1975, \mnras, 173, 729

\bibitem[{{Heggie} \& {Hut}(1993)}]{hegg93}
{Heggie}, D.~C., \& {Hut}, P. 1993, \apjs, 85, 347

\bibitem[{{Heggie} {et~al.}(1996){Heggie}, {Hut}, \& {McMillan}}]{hegg96}
{Heggie}, D.~C., {Hut}, P., \& {McMillan}, S.~L.~W. 1996, \apj, 467, 359

\bibitem[{{Hills}(1988)}]{hills88}
{Hills}, J.~G. 1988, \nat, 331, 687

\bibitem[{{Hills}(1991)}]{hil91}
---. 1991, \aj, 102, 704

\bibitem[{{Hobbs} \& {Nayakshin}(2009)}]{hob09}
{Hobbs}, A., \& {Nayakshin}, S. 2009, \mnras, 394, 191

\bibitem[{{Hopman}(2009)}]{hop09}
{Hopman}, C. 2009, \apj, 700, 1933

\bibitem[{{Hopman} \& {Alexander}(2006)}]{hop06}
{Hopman}, C., \& {Alexander}, T. 2006, \apj, 645, 1152

\bibitem[{{Hut}(1983)}]{hut83b}
{Hut}, P. 1983, \apj, 268, 342

\bibitem[{{Hut}(1993)}]{hut93}
---. 1993, \apj, 403, 256

\bibitem[{{Hut} \& {Bahcall}(1983)}]{hut83a}
{Hut}, P., \& {Bahcall}, J.~N. 1983, \apj, 268, 319

\bibitem[{{Kim} {et~al.}(2004){Kim}, {Figer}, \& {Morris}}]{kim2004}
{Kim}, S.~S., {Figer}, D.~F., \& {Morris}, M. 2004, \apjl, 607, L123

\bibitem[{{Kim} \& {Morris}(2003)}]{kim2003}
{Kim}, S.~S., \& {Morris}, M. 2003, \apj, 597, 312

\bibitem[{{Kocsis} \& {Tremaine}(2015)}]{koc15}
{Kocsis}, B., \& {Tremaine}, S. 2015, \mnras, 448, 3265

\bibitem[{Leigh {et~al.}(2016)Leigh, Antonini, Stone, Shara, \&
  Merritt}]{lei16a}
Leigh, N. W.~C., Antonini, F., Stone, N.~C., Shara, M.~M., \& Merritt, D. 2016,
  Monthly Notices of the Royal Astronomical Society, 463, 1605

\bibitem[{{Levin}(2007)}]{lev07}
{Levin}, Y. 2007, \mnras, 374, 515

\bibitem[{{L{\"o}ckmann} \& {Baumgardt}(2009)}]{loc09}
{L{\"o}ckmann}, U., \& {Baumgardt}, H. 2009, \mnras, 394, 1841

\bibitem[{{L{\"o}ckmann} {et~al.}(2009){L{\"o}ckmann}, {Baumgardt}, \&
  {Kroupa}}]{loc09b}
{L{\"o}ckmann}, U., {Baumgardt}, H., \& {Kroupa}, P. 2009, \mnras, 398, 429

\bibitem[{{Lu} {et~al.}(2013){Lu}, {Do}, {Ghez}, {Morris}, {Yelda}, \&
  {Matthews}}]{lu13}
{Lu}, J.~R., {Do}, T., {Ghez}, A.~M., {et~al.} 2013, \apj, 764, 155

\bibitem[{{Lu} {et~al.}(2009){Lu}, {Ghez}, {Hornstein}, {Morris}, {Becklin}, \&
  {Matthews}}]{lu09}
{Lu}, J.~R., {Ghez}, A.~M., {Hornstein}, S.~D., {et~al.} 2009, \apj, 690, 1463

\bibitem[{{Lucas} {et~al.}(2013){Lucas}, {Bonnell}, {Davies}, \&
  {Rice}}]{luc13}
{Lucas}, W.~E., {Bonnell}, I.~A., {Davies}, M.~B., \& {Rice}, W.~K.~M. 2013,
  \mnras, 433, 353

\bibitem[{{Madigan} {et~al.}(2011){Madigan}, {Hopman}, \& {Levin}}]{mad11}
{Madigan}, A.-M., {Hopman}, C., \& {Levin}, Y. 2011, \apj, 738, 99

\bibitem[{{Madigan} {et~al.}(2009){Madigan}, {Levin}, \& {Hopman}}]{madi09}
{Madigan}, A.-M., {Levin}, Y., \& {Hopman}, C. 2009, \apjl, 697, L44

\bibitem[{{Madigan} {et~al.}(2014){Madigan}, {Pfuhl}, {Levin}, {Gillessen},
  {Genzel}, \& {Perets}}]{madi14}
{Madigan}, A.-M., {Pfuhl}, O., {Levin}, Y., {et~al.} 2014, \apj, 784, 23

\bibitem[{{Mapelli} \& {Gualandris}(2016)}]{map16b}
{Mapelli}, M., \& {Gualandris}, A. 2016, 905, 205

\bibitem[{{Mapelli} {et~al.}(2008){Mapelli}, {Hayfield}, {Mayer}, \&
  {Wadsley}}]{map08}
{Mapelli}, M., {Hayfield}, T., {Mayer}, L., \& {Wadsley}, J. 2008, ArXiv
  e-prints, arXiv:0805.0185

\bibitem[{{Mapelli} \& {Trani}(2016)}]{map16a}
{Mapelli}, M., \& {Trani}, A.~A. 2016, \aap, 585, A161

\bibitem[{{McMillan} \& {Hut}(1996)}]{mcmill96}
{McMillan}, S.~L.~W., \& {Hut}, P. 1996, \apj, 467, 348

\bibitem[{{Merritt}(2010)}]{merr10}
{Merritt}, D. 2010, \apj, 718, 739

\bibitem[{{Mikkola} \& {Merritt}(2006)}]{mik06}
{Mikkola}, S., \& {Merritt}, D. 2006, \mnras, 372, 219

\bibitem[{{Mikkola} \& {Merritt}(2008)}]{mik08}
---. 2008, \aj, 135, 2398

\bibitem[{{Mikkola} \& {Tanikawa}(1999{\natexlab{a}})}]{mik99a}
{Mikkola}, S., \& {Tanikawa}, K. 1999{\natexlab{a}}, \mnras, 310, 745

\bibitem[{{Mikkola} \& {Tanikawa}(1999{\natexlab{b}})}]{mik99b}
---. 1999{\natexlab{b}}, Celestial Mechanics and Dynamical Astronomy, 74, 287

\bibitem[{{Murray-Clay} \& {Loeb}(2012)}]{mur12}
{Murray-Clay}, R.~A., \& {Loeb}, A. 2012, Nature Communications, 3, 1049

\bibitem[{{Naoz} {et~al.}(2018){Naoz}, {Ghez}, {Hees}, {Do}, {Witzel}, \&
  {Lu}}]{nao18}
{Naoz}, S., {Ghez}, A.~M., {Hees}, A., {et~al.} 2018, \apjl, 853, L24

\bibitem[{{Nayakshin} \& {Cuadra}(2005)}]{nay05}
{Nayakshin}, S., \& {Cuadra}, J. 2005, \aap, 437, 437

\bibitem[{{Nayakshin} {et~al.}(2007){Nayakshin}, {Cuadra}, \&
  {Springel}}]{nay07}
{Nayakshin}, S., {Cuadra}, J., \& {Springel}, V. 2007, \mnras, 379, 21

\bibitem[{{Nayakshin} \& {Zubovas}(2018)}]{naya18}
{Nayakshin}, S., \& {Zubovas}, K. 2018, \mnras, 478, L127

\bibitem[{{Paumard} {et~al.}(2006){Paumard}, {Genzel}, {Martins}, {Nayakshin},
  {Beloborodov}, {Levin}, {Trippe}, {Eisenhauer}, {Ott}, {Gillessen}, {Abuter},
  {Cuadra}, {Alexander}, \& {Sternberg}}]{pau06}
{Paumard}, T., {Genzel}, R., {Martins}, F., {et~al.} 2006, \apj, 643, 1011

\bibitem[{{Pelupessy} {et~al.}(2013){Pelupessy}, {van Elteren}, {de Vries},
  {McMillan}, {Drost}, \& {Portegies Zwart}}]{pel13}
{Pelupessy}, F.~I., {van Elteren}, A., {de Vries}, N., {et~al.} 2013, \aap,
  557, A84

\bibitem[{{Perets} \& {Alexander}(2008)}]{per08}
{Perets}, H.~B., \& {Alexander}, T. 2008, \apj, 677, 146

\bibitem[{{Perets} {et~al.}(2009){Perets}, {Gualandris}, {Kupi}, {Merritt}, \&
  {Alexander}}]{per09}
{Perets}, H.~B., {Gualandris}, A., {Kupi}, G., {Merritt}, D., \& {Alexander},
  T. 2009, \apj, 702, 884

\bibitem[{{Perets} {et~al.}(2007){Perets}, {Hopman}, \& {Alexander}}]{per07}
{Perets}, H.~B., {Hopman}, C., \& {Alexander}, T. 2007, \apj, 656, 709

\bibitem[{{Pfuhl} {et~al.}(2014){Pfuhl}, {Alexander}, {Gillessen}, {Martins},
  {Genzel}, {Eisenhauer}, {Fritz}, \& {Ott}}]{pfu14}
{Pfuhl}, O., {Alexander}, T., {Gillessen}, S., {et~al.} 2014, \apj, 782, 101

\bibitem[{{Plewa} {et~al.}(2017){Plewa}, {Gillessen}, {Pfuhl}, {Eisenhauer},
  {Genzel}, {Burkert}, {Dexter}, {Habibi}, {George}, {Ott}, {Waisberg}, \& {von
  Fellenberg}}]{ple17}
{Plewa}, P.~M., {Gillessen}, S., {Pfuhl}, O., {et~al.} 2017, \apj, 840, 50

\bibitem[{{Portegies Zwart} {et~al.}(2013){Portegies Zwart}, {McMillan}, {van
  Elteren}, {Pelupessy}, \& {de Vries}}]{por13}
{Portegies Zwart}, S., {McMillan}, S.~L.~W., {van Elteren}, E., {Pelupessy},
  I., \& {de Vries}, N. 2013, Computer Physics Communications, 184, 456

\bibitem[{{Portegies Zwart} {et~al.}(2009){Portegies Zwart}, {McMillan},
  {Harfst}, {Groen}, {Fujii}, {Nuall{\'a}in}, {Glebbeek}, {Heggie}, {Lombardi},
  {Hut}, {Angelou}, {Banerjee}, {Belkus}, {Fragos}, {Fregeau}, {Gaburov},
  {Izzard}, {Juri{\'c}}, {Justham}, {Sottoriva}, {Teuben}, {van Bever},
  {Yaron}, \& {Zemp}}]{por09}
{Portegies Zwart}, S., {McMillan}, S., {Harfst}, S., {et~al.} 2009, \na, 14,
  369

\bibitem[{{Portegies Zwart} {et~al.}(2003){Portegies Zwart}, {McMillan}, \&
  {Gerhard}}]{por2003}
{Portegies Zwart}, S.~F., {McMillan}, S.~L.~W., \& {Gerhard}, O. 2003, \apj,
  593, 352

\bibitem[{{Rauch} \& {Tremaine}(1996)}]{rau96}
{Rauch}, K.~P., \& {Tremaine}, S. 1996, NewA, 1, 149

\bibitem[{{Samsing} {et~al.}(2018){Samsing}, {Leigh}, \& {Trani}}]{sam18a}
{Samsing}, J., {Leigh}, N.~W.~C., \& {Trani}, A.~A. 2018, ArXiv e-prints,
  arXiv:1803.08215

\bibitem[{{Sana} {et~al.}(2012){Sana}, {de Mink}, {de Koter}, {Langer},
  {Evans}, {Gieles}, {Gosset}, {Izzard}, {Le Bouquin}, \& {Schneider}}]{san12}
{Sana}, H., {de Mink}, S.~E., {de Koter}, A., {et~al.} 2012, Science, 337, 444

\bibitem[{{Sch{\"o}del} {et~al.}(2007){Sch{\"o}del}, {Eckart}, {Alexander},
  {Merritt}, {Genzel}, {Sternberg}, {Meyer}, {Kul}, {Moultaka}, {Ott}, \&
  {Straubmeier}}]{sch07}
{Sch{\"o}del}, R., {Eckart}, A., {Alexander}, T., {et~al.} 2007, \aap, 469, 125

\bibitem[{{Shahzamanian} {et~al.}(2016){Shahzamanian}, {Eckart}, {Zaja{\v
  c}ek}, {Valencia-S.}, {Sabha}, {Moser}, {Parsa}, {Peissker}, \&
  {Straubmeier}}]{sha16}
{Shahzamanian}, B., {Eckart}, A., {Zaja{\v c}ek}, M., {et~al.} 2016, \aap, 593,
  A131

\bibitem[{{Spera} {et~al.}(2018){Spera}, {Mapelli}, {Giacobbo}, {Trani},
  {Bressan}, \& {Costa}}]{spera2018}
{Spera}, M., {Mapelli}, M., {Giacobbo}, N., {et~al.} 2018, ArXiv e-prints,
  arXiv:1809.04605

\bibitem[{{Stephan} {et~al.}(2016){Stephan}, {Naoz}, {Ghez}, {Witzel},
  {Sitarski}, {Do}, \& {Kocsis}}]{steph2016}
{Stephan}, A.~P., {Naoz}, S., {Ghez}, A.~M., {et~al.} 2016, \mnras, 460, 3494

\bibitem[{{Stoer} \& {Bulirsch}(1980)}]{sto80}
{Stoer}, J., \& {Bulirsch}, R. 1980

\bibitem[{{Sz{\"o}lgy{\'e}n} \& {Kocsis}(2018)}]{szo18}
{Sz{\"o}lgy{\'e}n}, {\'A}., \& {Kocsis}, B. 2018, ArXiv e-prints,
  arXiv:1803.07090

\bibitem[{{Toomre}(1964)}]{too64}
{Toomre}, A. 1964, \apj, 139, 1217

\bibitem[{{Trani} {et~al.}(2016{\natexlab{a}}){Trani}, {Mapelli}, {Bressan},
  {Pelupessy}, {van Elteren}, \& {Portegies Zwart}}]{tra16}
{Trani}, A.~A., {Mapelli}, M., {Bressan}, A., {et~al.} 2016{\natexlab{a}},
  \apj, 818, 29

\bibitem[{{Trani} {et~al.}(2016{\natexlab{b}}){Trani}, {Mapelli}, {Spera}, \&
  {Bressan}}]{tra16b}
{Trani}, A.~A., {Mapelli}, M., {Spera}, M., \& {Bressan}, A.
  2016{\natexlab{b}}, \apj, 831, 61

\bibitem[{{{\v S}ubr} {et~al.}(2009){{\v S}ubr}, {Schovancov{\'a}}, \&
  {Kroupa}}]{sub09}
{{\v S}ubr}, L.~., {Schovancov{\'a}}, J., \& {Kroupa}, P. 2009, \aap, 496, 695

\bibitem[{{{\v S}ubr} \& {Haas}(2014)}]{sub14}
{{\v S}ubr}, L., \& {Haas}, J. 2014, \apj, 786, 121

\bibitem[{{{\v S}ubr} \& {Haas}(2016)}]{subr16}
---. 2016, \apj, 828, 1

\bibitem[{{Witzel} {et~al.}(2014){Witzel}, {Ghez}, {Morris}, {Sitarski},
  {Boehle}, {Naoz}, {Campbell}, {Becklin}, {Canalizo}, {Chappell}, {Do}, {Lu},
  {Matthews}, {Meyer}, {Stockton}, {Wizinowich}, \& {Yelda}}]{wit14}
{Witzel}, G., {Ghez}, A.~M., {Morris}, M.~R., {et~al.} 2014, \apjl, 796, L8

\bibitem[{{Yelda} {et~al.}(2014){Yelda}, {Ghez}, {Lu}, {Do}, {Meyer}, {Morris},
  \& {Matthews}}]{yel14}
{Yelda}, S., {Ghez}, A.~M., {Lu}, J.~R., {et~al.} 2014, \apj, 783, 131

\end{thebibliography}

\appendix

\section{Repeated encounters}\label{sec:repeat}
The surviving binaries can also undergo multiple encounters, if the encounter rate is high enough. This does not alter only the orbital parameters of the inner binary, but also the orbital parameter of the binary around the SMBH. In this section we investigate whether these repeated encounters can drive the binaries to migrate in the $a$--$e$ space, until they are finally ionized by an encounter. In principle, this might result in the production of stars in highly eccentric orbits well below the inner edge of the original disk at $0.03 \pc$. 

\begin{figure}[htb]
    \centering
	\includegraphics[width=\linewidth]{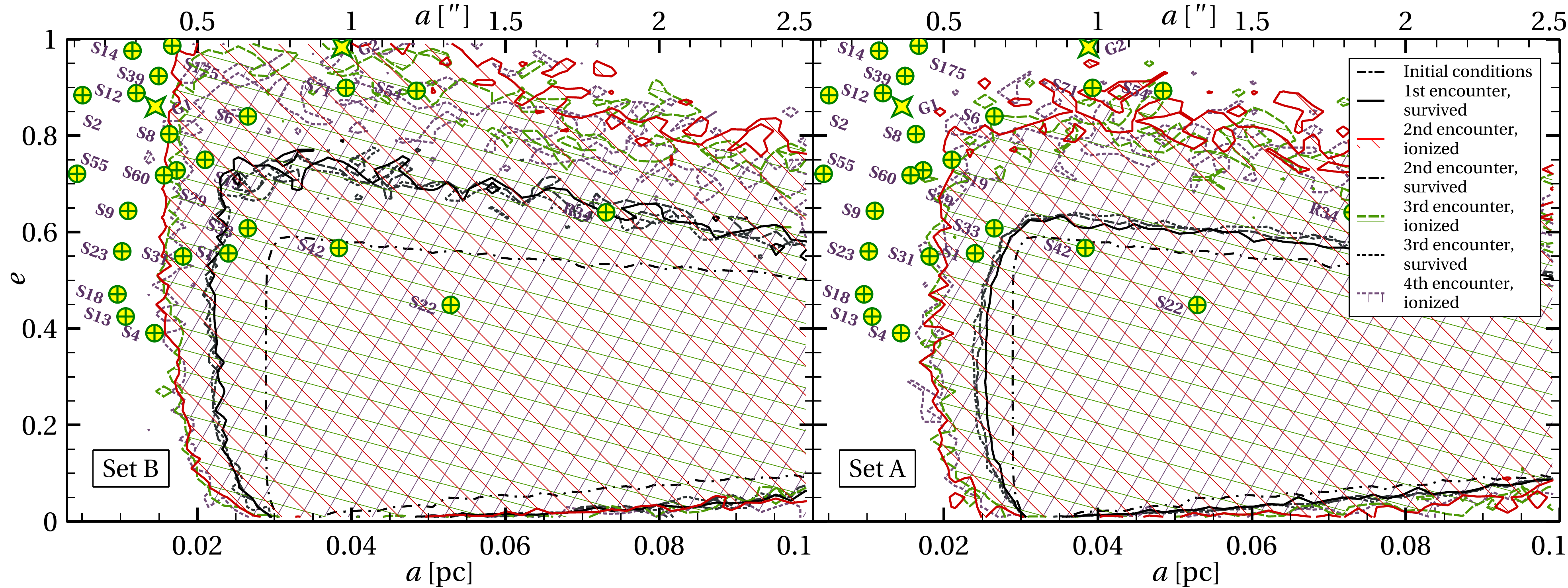}
	\caption{ 
			Semimajor axis-eccentricity map of ionized binary components and surviving binaries for multiple encounters in set~B (left) and A (right).
			Yellow circles with green cross: S-stars \citep[from][]{gil17}. Yellow stars with green contour: G1 and G2 objects. Black contours: binaries initial conditions derived from the first (solid), second (dashed) and third (dotted) encounter. Red solid contour with backward diagonals fill: ionized binary components in the second encounter. Green dashed contour with backward diagonal fill: ionized binary components in the third encounter. Purple dotted contour with forward diagonal fill: ionized binary components in the fourth encounter.
	}
	\label{fig:repeat}
\end{figure}

We take the orbital properties of the binaries that survive the encounter in set A and B, and use them as initial conditions for a new encounter. We repeat this procedure in order to simulate 3 encounters after the first.
Figure~\ref{fig:repeat} shows the distribution of the surviving binaries and corresponding ionized stars after the multiple encounters. Because the binaries are, on average, more massive then the stellar black hole, the binaries do not migrate significantly in the $a$--$e$ space with respect to the initial conditions. As a consequence, the distribution of stars ionized after multiple encounters does not change very much with respect to the distribution shown in Figure~\ref{fig:aemap}. 

\section{Simulations with different black holes masses}\label{sec:m10}

\begin{figure}[htb]
	\centering
	\includegraphics[width=1\linewidth]{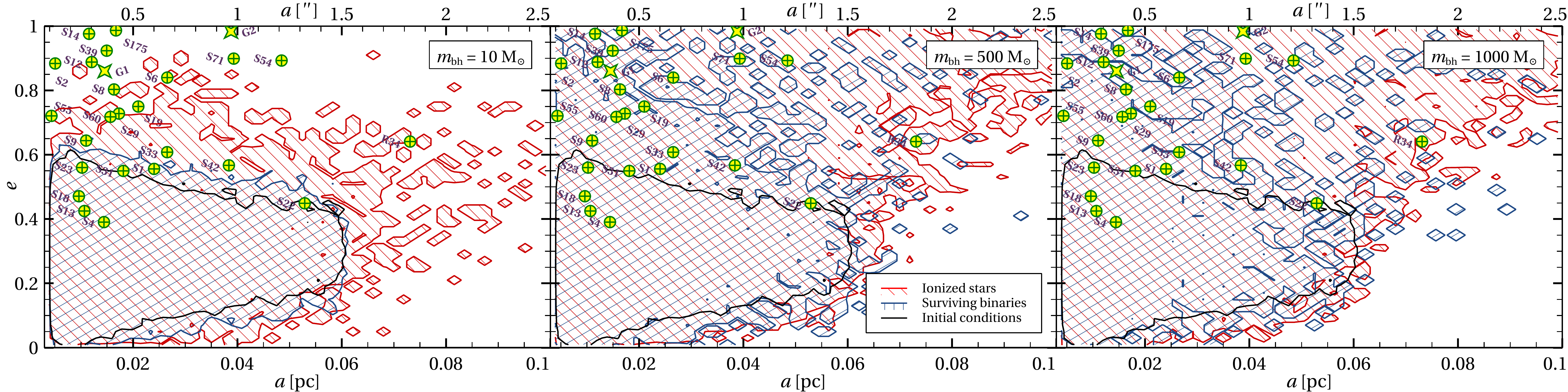}
	\caption{ 
		Semimajor axis-eccentricity map of ionized binary components in set~B, for encounters with a stellar black hole mass of $10\msun$ (red contour with backward diagonal fill) and $30 \msun$ (blue contour with forward diagonal fill). Black contour: initial conditions.
		Yellow circles with green cross: S-stars \citep[from][]{gil17}. Yellow stars with green contour: G1 and G2 objects.}
	\label{fig:m10}
\end{figure}

{ 
	To check if also black holes with smaller masses can result in the ionization of B-type binaries, we have three additional sets of simulations using the same setup of set~\Bex but setting the mass of black hole $10\msun$, $500\msun$ and $1000\msun$. In Figure~\ref{fig:m10} we show the $a$--$e$ distribution of the ionized stars and surviving binaries for this supplementary sets.

	As expected, the ionized stars reach a lower eccentricity when the stellar black hole mass is smaller. Nonetheless, the achieved eccentricity is still high enough to match the orbital properties of several S-stars with $e\lesssim 0.8$. Conversely, for $m>500\msun$ (i.e. intermediate mass black holes, IMBHs), more stars get ionized and kicked into highly eccentric orbits.
	
	Interestingly, for $m_\mathbf{bh} = 500$ and $1000 \msun$, also binary stars are scattered into highly eccentric orbit. 
	Current observations do not rule out that some of the S-stars may be in fact binaries \citep{chu18}.
	Furthermore, the binary merger scenario for the origin of G2 requires the presence of binaries in highly eccentric orbits about the SMBH \citep[][see also \citealt{steph2016}]{wit14}.
	
	The formation of binary S-stars via encounters with IMBHs will be investigate in more detail in our forthcoming work.
	
}

\end{document}